# On-chip erbium-doped lithium niobate waveguide amplifiers

Qiang Luo[1], Chen Yang[1], Zhenzhong Hao[1], Ru Zhang[1], Dahuai Zheng[1], Fang Bo[1,2,3,*], Yongfa Kong[1,*], Guoquan Zhang[1,*], Jingjun Xu[1,*]

[1] MOE Key Laboratory of Weak-Light Nonlinear Photonics, TEDA Institute of Applied Physics and School of Physics, Nankai University, Tianjin 300457, China

[2] Collaborative Innovation Center of Extreme Optics, Shanxi University, Taiyuan 030006, Shanxi, China

[3] Collaborative Innovation Center of Light Manipulations and Applications, Shandong Normal University, Jinan 250358, China

*Corresponding author: bofang@nankai.edu.cn; kongyf@nankai.edu.cn; zhanggq@nankai.edu.cn; jjxu@nankai.edu.cn

Lithium niobate on insulator (LNOI), as an emerging and promising optical integration platform, faces shortages of on-chip active devices including lasers and amplifiers. Here, we report the fabrication on-chip erbium-doped LNOI waveguide amplifiers based on electron beam lithography and inductively coupled plasma reactive ion etching. A net internal gain of ~30 dB/cm in communication band was achieved in the fabricated waveguide amplifiers under the pump of a 974-nm continuous laser. This work develops new active devices on LNOI and will promote the development of LNOI integrated photonics.

## 1. Introduction

As an excellent optical crystal material, lithium niobate (LiNbO$_3$ or LN) is considered as one of the optical integration platform by virtue of small absorption coefficient (0.02 cm$^{-1}$ at 1064 nm), wide transparent window (0.35-5 μm), high nonlinear effect ($d_{33}$ = -41.7 pm/V at 1.058 μm), strong electro-optic effect ($r_{33}$ = 32.2 pm/V at 1.547 μm) and acousto-optic effect. Traditionally, integrated optical devices were produced based on Ti-diffusion or proton-exchanged waveguides with weak refractive index contrast (~0.1) and large waveguide width (~10 μm), which hinders the development of integrated photonics on LN platform. Fortunately, with the development of LNOI and the advance of corresponding micro-/nano-processing technologies, wire and ridge waveguides with high refractive index contrast, small footprint and low loss were demonstrated[1, 2]. The propagating loss of the LNOI waveguide with micrometer scale has been reduced to as low as 0.027 dB/cm[3, 4]. LNOI microresonators with quality factors higher than 10$^8$ were also reported very recently, whose loss approaches the material absorption limit[5]. In recent years, passive integrated optical devices such as frequency converters[6-13], electro-optic modulators[14, 15], optical frequency combs[16-18], and spectrometer[19] have been developed on the LNOI platform.

Another important part of integrated optical systems is the active LNOI devices, such as lasers and amplifiers. It is known that LN is an indirect band gap material emitting light inefficiently. The luminescence of LN can be achieved with the help of doping rare earth ions. Recently, some researchers first prepared waveguide and microring structures on LNOI using electron beam lithography (EBL) and dry etching processes, then doped them with erbium and ytterbium ions by ion implantation followed by thermal annealing at ~500℃ to repair the lattice damage. Based on these rare earth ions doped LNOI devices photoluminescence experiments were carried out[20, 21]. However, limited by the low doping concentration and the inhomogeneous distribution of rare earth ions, no laser radiation or light amplification is observed. Ingeniously, a good solution was conducted that first doped LN crystals during the growth process, and then use ion-cut technology to prepare rare earth-doped LN films with high doping concentration and uniform ion distribution. Currently, thulium and erbium doped LNOI platforms have been reported based on this method[22-25] and 1550-nm band erbium-doped LNOI micro-disk lasers were achieved very recently[23-25].

The on-chip LNOI amplifier, as a fundamental optical component of active LNOI devices, however, is rarely reported. Here, we present the fabrication of LNOI waveguide amplifiers using EBL and inductively coupled plasma reactive ion etching (ICP-RIE) process on homemade erbium-doped LNOI wafer. Under the 980-nm band pump, a net internal gain of ~15 dB at 1531.5 nm was realized on a ~5 mm long waveguide structure. The dynamic behavior of gain from linear to saturation was observed in the cases of both fixed signal with increasing pump power and fixed pump power with increasing signal power.

## 2. Fabrication of erbium-doped LNOI waveguide

Fabrication of erbium-doped waveguide amplifiers starts from an erbium-doped X-cut LNOI wafer with a doping concentration of 0.1 mol%. The wafer is composed of a 600-nm-thick erbium-doped LN film that was sliced from a homemade LN crystal wafer, a 2-μm-thick of silicon dioxide ($SiO_2$) buffer layer and a 500-μm-thick silicon substrate. The preparation process is schematically illustrated in Fig. 1(a), which is mainly divided into four steps. Firstly, a layer of hydrogen silsesquioxane (HSQ) resist was spin-coated on the erbium-doped LN film. Subsequently, the patterns of waveguide amplifiers were defined by EBL. Then, $Ar^+$ plasma etching was carried out to transfer the mask patterns into the erbium-doped LN film, resulting in ridge waveguides with a 280-nm etching depth and a 60° wedge angle. Finally, the chip was immersed in HF solution for 5 min to remove residual resist mask. Lastly, the facets of the optical waveguides were mechanically processed for efficient fiber to facet coupling. Figure 1(b) shows the optical micrograph of an array of fabricated erbium-doped LNOI waveguides. The final length of the straight waveguide is about 5 mm. Ridge LNOI waveguides in micrometer scale have a high refractive index contrast leading to strong light field confinements in both 980-nm band and in 1550-nm band. Based on the structure parameters of the fabricated waveguides with a top width of 1.4 μm, the field distributions and effective refractive indices of waveguide eigenmodes can be numerically calculated. The mode distributions of fundamental transverse electric modes at pump (~974.3 nm) and signal (~1531.5 nm) wavelengths were calculated and are shown in Fig. 1(c) as examples.

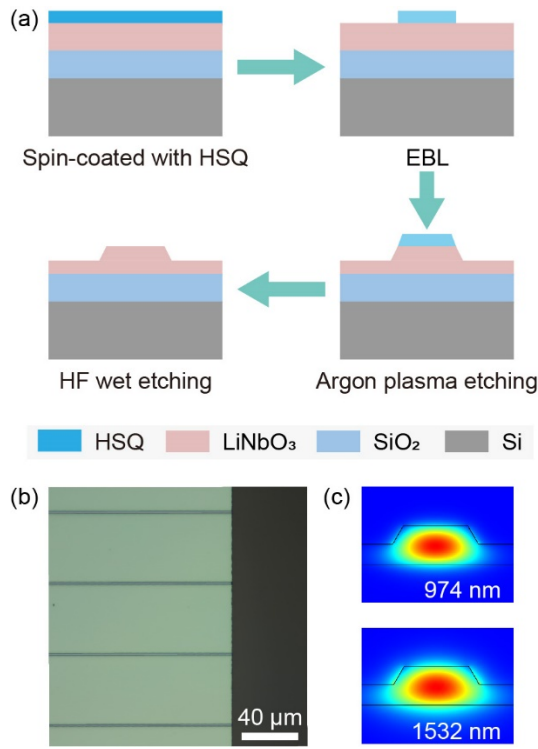

**Fig. 1.** (a) The schematic fabrication processes of erbium-doped LNOI waveguides. (b) Optical micrograph of fabricated erbium-doped LNOI waveguides. (c) Simulation result of mode distributions regarding fundamental transverse electric modes at wavelengths of 974 nm (top) and 1532 nm (bottom).

## 3. Characterizations of erbium-doped LNOI waveguide amplifiers

Respecting the higher absorption coefficient of erbium ions in 980-nm band compared with that in 1480-nm band[26], a continuous laser working at 974.3 nm was selected as the pump to investigate the optical amplification performance of the erbium-doped LNOI waveguides. A tunable narrow-band laser operating in 1550-nm band served as the signal laser. Figure 2 schematically illustrates the experimental setup for the characterization of the LNOI amplifier. After passing through a variable optical attenuator (VOA), an optical coupler (OC) and polarization controller (PC), respectively, the pump and signal lasers were combined using a wavelength division multiplexer (WDM) and then launched into LNOI waveguide via a lensed fiber. At the same time, the pump/signal power in the optical path was monitored by sending the light from the second port of the corresponding OC to a power meter (PM). While propagating in the on-chip waveguide, pump is absorbed by erbium ions and converted into gain to realize signal amplification under the signal seeds light induction (stimulated emission). Similarly, both the amplified signal and residual pump light output from the chip were collected by a second lensed fiber and sent to an optical spectrum analyzer to detect the 1550-nm band signal. It is worth noting that under higher pump power levels, a strong green up-conversion fluorescence along the waveguide was generated as shown in the device photography in Fig. 2.

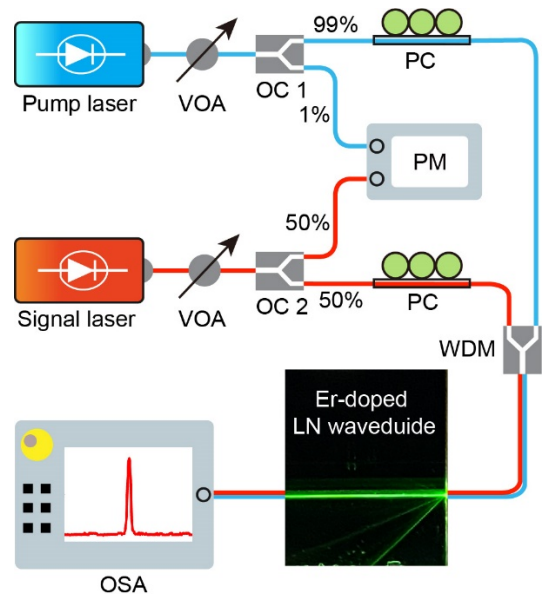

**Fig. 2.** Experimental setup for gain characterization in erbium-doped LNOI waveguide amplifiers. VOA: variable optical attenuator; OC: optical coupler; PM: power meter; PC: polarization controller; WDM: wavelength division multiplexer; OSA: optical spectrum analyzer. The photograph of erbium-doped LNOI chip clearly shows the generated green fluorescence propagating along the straight waveguide.

To calibrate the net internal gain of erbium-doped LNOI waveguide amplifiers, we first characterized the optical propagation losses for both the pump and the signal of chip waveguides using whispering-gallery-resonator-loss measurements based on an erbium-doped microring resonator coupling with waveguide structure on the same chip. As shown in Fig. 3, through fitting the resonance spectrum by a Lorentz function, the load $Q_L$ at 1531.5 nm and 974.5 nm resonance mode were derived as $9.78 \times 10^4$ and $2.42 \times 10^5$, respectively. According to coupling state (1550-nm band over-coupling and 980-nm band under-coupling) inferred by gap between the microring and the waveguide, the intrinsic quality $Q_i$ were obtained as $5.0 \times 10^5$ (1531.5 nm) and $2.6 \times 10^5$ (974.5 nm), respectively. Finally, the propagation loss coefficient $\alpha$ was estimate based on

$$\alpha = \frac{2\pi n_{eff}}{\lambda Q_i}, \quad (1)$$

where the $n_{eff} = \lambda^2/(2\pi R \cdot FSR)$ is the effective refractive at the relevant wavelength, $R$ and $FSR$ denoting the microring radius (100 μm) and free spectra range, respectively. The calculated propagation losses for the signal and pump are 0.83 dB/cm and 2.5 dB/cm, respectively. The propagation loss is mainly composed of the scattering loss of the waveguide side wall and the absorption loss from erbium ions. Assuming that both ends are identical, we estimated the single-end-facet coupling losses as 6.76 dB at 1531.5 nm and 6.35 dB at 974.5 nm by considering the fiber-to-fiber insertion loss and chip propagation loss. Based on the above calibration results, the pump and signal power are referred to on-chip power in this paper.

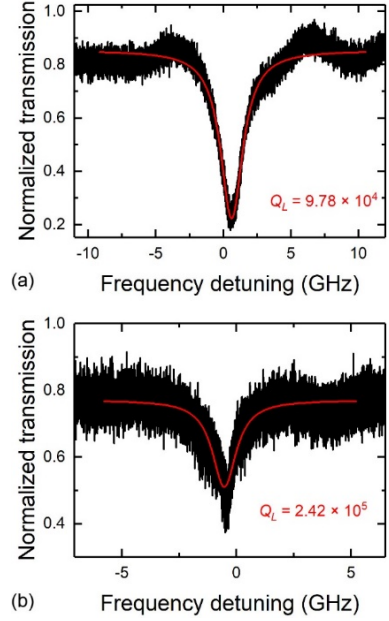

**Fig. 3.** Optical transmission spectra of Er-doped LNOI microring resonators on the same chip in 1550-nm band (a) and 980-nm band (b). The lorentz-fit (red line) showing $9.78 \times 10^4$ and $2.42 \times 10^5$ loaded quality factors near 1531.5 nm and 974.5 nm, respectively.

The internal net gain of the erbium-doped LNOI waveguide amplifier was defined as

$$g = 10\log_{10}\frac{P_{on}}{P_{off}} - \alpha L \quad (2)$$

where the $P_{on}$ and $P_{off}$ are the output signal power in pump-on and pump-off cases, $\alpha$ and $L$ represent the propagation loss coeffcient at signal wavelength and waveguide length, respectively. Here, $\alpha L$ is estimated as 0.42 dB. The net internal gain under pump power from 0 to 64 mW is shown in Fig. 4(a) with fixed signal power ~5 nW at 1531.5 nm. At low pump power, the net internal gain increases rapidly with the increasing of pump power, which belongs to the small signal amplification stage. Obviously, the internal loss is enough to compensate with pump power at ~1 mW. When the pump power increases to 60 mW, the net internal gain tends to be saturated. The evolution of the measured signal spectrum with the increase of pump power (0, 0.10 mW, 7.35 mW, 16.19 mW, 32.31 mW, 64.02 mW) is shown in Fig. 4(b). The maximum net internal gain of ~5.5 dB was obtained in our device within a waveguide length of ~5 mm at higher pump power ~64 mW, which corresponding to a net internal gain of 11 dB/cm. Such a net gain could be further improved by increasing the doping concentration of erbium ions. Compared with traditional Er:Ti:LiNbO3 waveguide amplifier with 2 – 3 dB/cm gain, the net internal gain has been significantly improved[27-30]. This is mainly due to the increase of optical power density for both pump and signal modes and the spatial overlap with erbium ions.

Moreover, we characterized the gain dependence of erbium-doped LNOI waveguide amplifier on the signal power. As shown in Fig. 4(c), at low signal power, the net internal gain decreases linearly with the increasing signal power for a fixed pump power of ~23 mW, corresponding to the small signal gain state (-65 dBm to -50 dBm). As the signal power continues to increase, gain saturation is observed. The maximum gain is ~15 dB (~30 dB/cm) with the signal power of ~65 dBm, and there is still ~0.13 dB amplification at ~34 dBm signal power. As shown in the amplified spontaneous emission (ASE) spectrum of the erbium-doped LNOI waveguide with signal power of 0.57 mW (Fig. 4(d)), although the optimal working wavelength is around ~1531.5 nm, the amplifier works in a wide range of communication band.

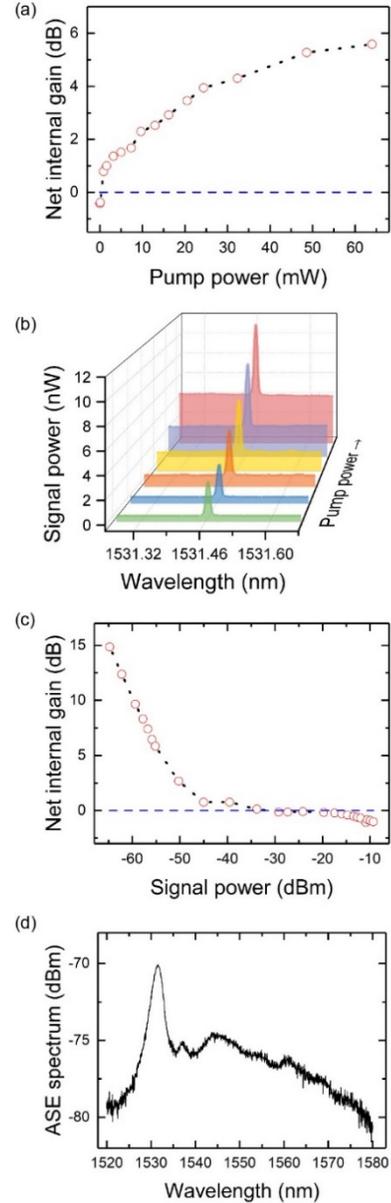

**Fig. 4.** Gain characterization in erbium-doped LNOI waveguide amplifiers. (a) The dependence of net internal gain on pump power at fixed signal power ~5 nW. (b) Measured signal spectra at ~1531.47 nm with an increasing pump powers, pump power: 0, 0.10 mW, 7.35 mW, 16.19 mW, 32.31 mW, 64.02 mW. (c) The net internal gain as a function of increasing signal power at fixed pump power ~23 mW. (d) The amplified spontaneous emission (ASE) spectrum of the erbium-doped LNOI waveguide at signal power 0.57 mW

## 4. Conclusions

In summary, we fabricated on-chip erbium-doped LNOI waveguide amplifiers using EBL and ICP-RIE processes. Under 980-nm band pump, the communication band amplifiers were demonstrated with maximum net internal gain of ~15 dB achieved in a 5-mm-long chip. Compared with the previous Er:Ti:LiNbO$_3$ bulk waveguide amplifier, the amplification performance has been greatly improved. The amplifier could be compatible with passive LNOI devices by selectively doping LN wafer via erbium diffusion during the fabrication of LNOI wafer, which may significantly promote the development of LNOI on-chip integrated LNOI optics.

During the preparation of this article, we noticed that two erbium-doped waveguide amplifiers works were posted on arXiv[31, 32], in which the authors had reported similar amplification performance. The differences of our works are as follows: (1) Our preparation process is more simplified, without Cr film deposition and CMP polishing steps. (2) Our doping concentration of erbium ions is relatively lower, which makes our amplifier have relatively high net gain per unit length for weak signal (~30 dB at signal power of ~-65 dBm), so it may have more advantages in weak signal amplification.

## Acknowledgement
This work was supported by the National Key Research and Development Program of China (Grant No. 2019YFA0705000), the National Natural Science Foundation of China (Grant Nos. 12034010, 11734009, 92050111, 92050114, 12074199, 12004197, and 11774182), the Higher Education Discipline Innovation Project (Grant No. B07013) and the Program for Changjiang Scholars and Innovative Research Team in University (PCSIRT) (Grant No. IRT_13R29).